# The adaptability of physiological systems optimizes performance: new directions in augmentation.

Bradly Alicea
**MIND Lab, Michigan State University, East Lansing, MI 48824**

Keywords: Learning and Memory, Complex Systems, Human-machine Systems, Augmented Cognition

**Abstract.** This paper contributes to the human-machine interface community in two ways: as a critique of the closed-loop AC (augmented cognition) approach, and as a way to introduce concepts from complex systems and systems physiology into the field. Of particular relevance is a comparison of the inverted-U (or Gaussian) model of optimal performance and multidimensional fitness landscape model. Hypothetical examples will be given from human physiology and learning and memory. In particular, a four-step model will be introduced that is proposed as a better means to characterize multivariate systems during behavioral processes with complex dynamics such as learning. Finally, the alternate approach presented herein is considered as a preferable design alternate in human-machine systems. It is within this context that future directions are discussed.

## 1.0 Introduction

This paper is an attempt to develop more complex modes of mitigation strategy for Augmented Cognition (AC) applications. A secondary goal will be to introduce the concept of adaptability to the science of human performance. To do this, I will review the concepts of fitness and fitness landscapes as they relate to human physiological systems and life-history processes such as plasticity, neuroprotection, and aging. Adaptive changes across human life-history involves several processes which, while fixed in their genetic structure, can express much hidden or latent variation that may affect how both physiological and cognitive processes unfold (Shostak, 2006). This affects our conception of performance augmentation, from the optimization of a simple function with a linear response (Schmorrow and Stanney, 2008) to the guidance of complex, nonlinear phenomenon towards optimal states (see Figure 1).

## 1.1 Example of simple model of mitigation

The canonical example of a mitigation strategy involves the Yerkes-Dodson curve (Schmorrow, 2005). The Yerkes-Dodson curve is an inverted-U shape function that describes the general state of arousal in relation to performance. The inverted-U is a classic prediction of the relationship between a physiological indicator and behavioral performance, although it is not the only or even most predictive model of this relationship (Kvetnansky et.al, 1991, Page 983). In the Yerkes-Dodson inverted-U model (referred to herein as the Gaussian model), moderate states of arousal are considered to be optimal for performance settings such as driving and operating machinery.

A mitigation strategy is simply a technological augmentation applied to physiological systems so that the behavioral output is pushed back into the optimal range. An example from the Yerkes-Dodson curve (top frame of Figure 1) shows this relationship nicely: when the level of arousal reaches a lower-bound threshold (i.e. the operator is getting drowsy), the augmentation is implemented and the measured level of arousal recovers.

The mitigation acts as a complex feedback loop in that one form of augmentation must be applied as the arousal measure reaches a lower-bound but another form of augmentation must be applied as that same measure reaches an upper-bound.

**1.2 Critique of a Gaussian model of augmentation**

The top frame of Figure 1 shows an example of the classical (or Gaussian) model for determining a mitigation strategy. The method is simple: augment behavior whenever some gauge of performance is below or above a certain value. However, there are numerous problems with using such a model over the course of training and more generally during long-term interaction with a system.

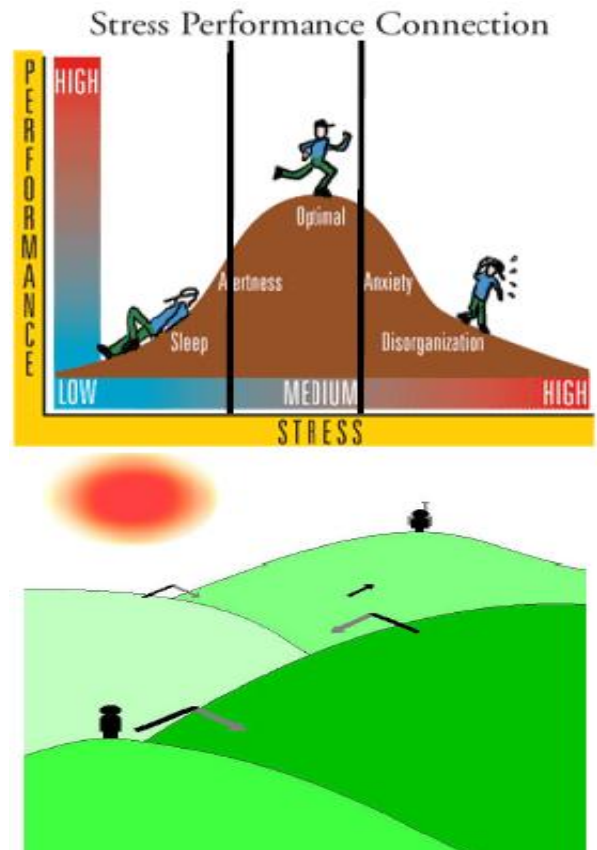

**Figure 1. Top frame: classical view of a mitigation strategy (optimization of a simple function – see black bars). Bottom frame: proposed view of a mitigation strategy (guidance of a complex nonlinear phenomenon towards optimal states).**

*1.2.1 Shortcomings of Gaussian response function for complex behavior.* It has been pointed out by Lacey (in Cacioppo et.al, 2000) that the Gaussian model which reliably characterizes phenomena such as the fight-or-flight response may be inappropriate for more complex behaviors. In terms of relatively simple responses such as arousal, there are three known indicators of the same phenomenon, each with an independent profile. Furthermore, response characteristics are known to be specific to context and physiological subsystem (Cacioppo et.al, 2000, Pages 53-54). Therefore, indicators for separate subsystems that produce a single behavior may evolve in different directions over the course of real-time behavior. These divergent trajectories may affect the inverted-U in ways that are a product of both simple habituation and more complex dynamics.

## 1.3 Augmentation from a microevolutionary perspective

From an evolutionary perspective, constraining the set of behavioral and physiological outputs to a finite range constitutes artificial selection. Artificial selection is a commonly observed in the form of domestication, but has also been used extensively as an experimental tool across generations to demonstrate how natural selection can produce novel, specialized traits (Garland, 2003). Within a life span, artificial selection can act to either expand or constrain the adaptive response of the physiological systems involved in determining performance level by selecting for certain physiological responses over others.

While these changes are not heritable, they can potentially affect gene expression, epigenetic, and hormonal responses. This can have both positive and deleterious effects on performance. One example from selection applied across multiple generations is the evolution of senescence (e.g. aging). In this case, natural selection for specific alleles (e.g. genetic variants) can postpone deleterious gene expression during the life span, allowing people to age slower and live longer (Bell, 1997; Shostak, 2006).

In the case of more immediately expressed traits and processes such as arousal and stress, the lack of expressed variation may make these states and their associated metabolic processes inaccessible over time. While this seems to be a good thing for suboptimal performance states, this is also analogous to how muscles atrophy during periods of extended disuse. That is, the experience of stress and suboptimal states on a regular basis keeps the system running in ways that have a positive overall on the individual.

*1.3.1 Hysteresis of the response curve.* A specific example of how physiological sensitization and selection relate to the lack of variability present in systems utilizing a simple form of mitigation can be seen in the hysteresis[1] of the performance response curve. While the path to optimality may often involve temporarily reduced levels of performance, the effects of augmentation on the immediate physiological response may also be a factor in whether or not a given mitigation strategy will actually work. Systematic changes in the response curve due to adaptive mechanisms can be referred to as hysteresis. Hysteresis is directly affects by the degree of mitigation, and determines the amount of optimal behavior available to the physiological system to drive further adaptation (see Figure 2).

## 1.4 Introduction to complex model of mitigation

The complex model of mitigation is based on concepts from evolutionary and complexity theory, and treats environmental fitness and natural selection as variables that can vary and be measured within a single generation. When variability in fitness and selection are measured across generations, this is typically related to changes in gene frequencies and the fixation of traits by neutral demographic processes. However, there are multiple time scales at which natural selection operates, and given the appropriate target phenomena such mechanisms can explain a lot.

---

[1] Hysteresis can be defined as the influence of a system's history on its current state. In the context of a mathematical function, this results in a change in function shape.

*1.4.1 Fitness and selection as variables in mitigation strategy.* Within a single life-history, natural selection and fitness relate more to the initiation of gene expression patterns and standing individual variation that has gotten there through processes operating across generations. As was demonstrated in the case of artificial selection, fitness and selection can be measured and applied to adaptive processes such as neuroplasticity and neuroprotection[2]. For purposes of this paper, we can define two coefficients that act as variables in the complex mitigation model. This first of these is the fitness coefficient ($F_c$). Traditional examples of the fitness coefficient are determined by individual behavioral performance, fecundity, or survival compared to all other members of a population[3]. Conceptually, fitness captures how well an organism meets the challenges of a variable environment. The second of these is the selection coefficient ($S_c$). Traditional examples of the selection coefficient are determined by the degree of environmental challenge presented to an organism[4]. Conceptually, environmental selection determines in part how rugged fitness landscape is in a given setting.

## 2.0 Introduction to Fitness

In evolutionary biology, fitness is a central concept that assesses the relative competitiveness of one unit in comparison with all others in the population. The term unit is used in the definition because fitness can be assessed at multiple scales of analysis, from molecular biology to social groups. Fitness can also be used to assess the output of organismal systems (e.g. behavior and physiological indicators), which incorporates the features of multiple scales of analysis.

In the case of assessing performance, such a measure allows us to better understand the trends in this output, and how it might allow for adaptation. Sensory stimulation and stress can trigger changes in anatomical and molecular signaling pathways that can affect performance in ways not immediately obvious. Given the right degree of selection, this adaptation may lead to even higher levels of performance. In addition, different individuals may exhibit differential solutions to a specific environmental challenge. Therefore, complex mitigation can cater to highly contextual scenarios. For example, being "fit" in one setting does not always guarantee universal fitness.

## 2.1 Fitness landscapes

Fitness landscapes are an old concept applied to a diverse range of problems over the years. Sewall Wright originally applied this concept to the study of population genetics (Wright, 1932). Later, Kaufmann and Wagner applied them to the study of complexity in natural systems: Kaufmann to NK-boolean nets (see Kauffman, 1993) and Wagner to the evolvability of genotypes and phenotypes (see Wagner, 2007).

---

[2] for definition of neuroprotection, see Gidday, 2006 and Stenzel-Poore, 2004.

[3] the relative fitness coefficient can be calculated using the following equation: $\mathbf{F_c = (X_n - X_{min} / X_{max} - X_{min})}$. F is the fitness coefficient, $X_n$ is the measurement of a single variable for a single subject, $X_{min}$ is the lower bound value for a given variable in a sample, and $X_{max}$ is the upper bound value for a given variable in a sample.

[4] the selection coefficient can be calculated using the following equation: $\mathbf{S_c = N_s - D}$. $S_c$ is the selection coefficient, $C_s$ is the control stimulus, and D is degradation, characterized quantitatively by the degree of masking in a visual or auditory channel.

Phenomenologically, a fitness landscape is an *n*-dimensional space. This is generally visualized in theoretical examples as a 3-D space that is similar to a trend surface. The X and Y-axes of this surface represents the matrix position of all possible environmental challenges, while the Z-axis is the measure of fitness required for particular X, Y location in this matrix (see Figure 3). In the *n*-dimensional case, the fitness and/or environmental representation can represent up to *n*-1 dimensions of the space, as both fitness and the environment can both be high-dimensional constructs. In the 3-D case, the matrix of X, Y environmental setting approximations is a combinatoric measure of context (see Figure 3). This allows us to approximate not only the effects of known environmental conditions, but the effects of hypothetical conditions as well.

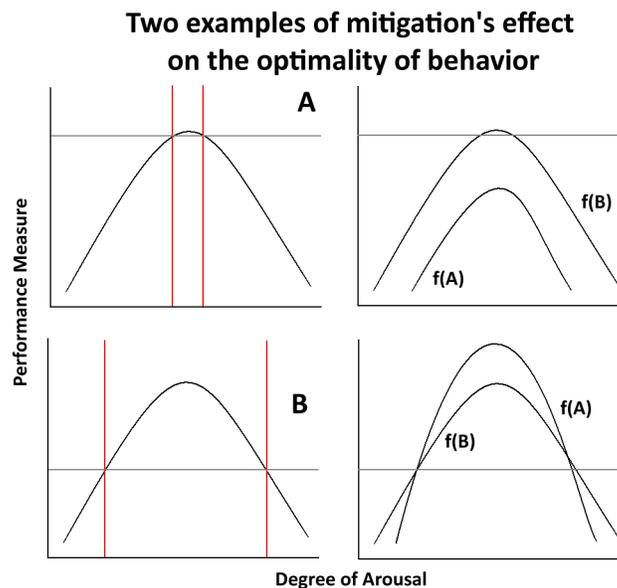

**Figure 2. Two examples of mitigation's effect on performance output using the Gaussian model. This example assumes the Yerkes-Dodson model of arousal. The red lines are upper- and lower-bound thresholds, and everything above the grey horizontal lines represents "optimal" behavior. In frames A, a smaller range of optimal behavior (left) results in downward hysteresis of the response function after augmentation (right). In frames B, a larger range of optimal behavior (left) results in upward hysteresis of the response function after augmentation (right).**

*2.1.1 Significance of landscape topology.* In the 3-D case, a fitness landscape surface has a variable topology which ranges from "smooth" to "rugged" (see Figure 4). "Smooth" landscapes are essentially an isomorphic mapping between individual fitness and environmental challenges, while "rugged" landscapes represent a mismatch between individual fitness and environmental challenges.

This can be demonstrated by considering the effects of environment on an organism in each case. For smooth landscapes, the organism can be placed in any environmental setting and no adaptation is required on the part of the physiology. For rugged landscapes, the organism can still be put in any environmental setting, but in some cases significant adaptation will be required on the part of the physiology.

In general, the amount of adaptation required and degree of adaptation possible are independent processes. However, the relationship between these processes can determine degree of mismatch or ruggedness. This is especially true as they relate to the selection and fitness coefficients (e.g. $S_c / F_c > 0$).

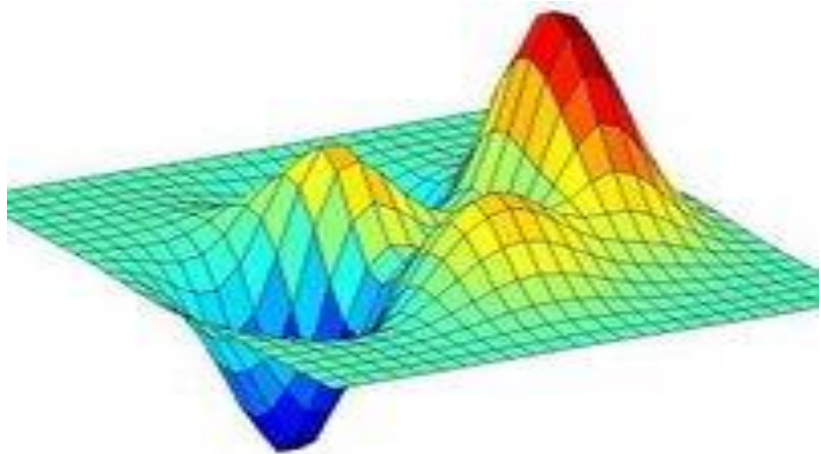

**Figure 3: A simple example of a fitness landscape.**

## 2.2 Mathematical representation of adaptive states

Each state in the complex model can be defined by looking at the interaction between the selection and fitness coefficients. The phase portrait shown in Figure 5 demonstrates the static relationship between these variables and under what conditions these states are achieved.

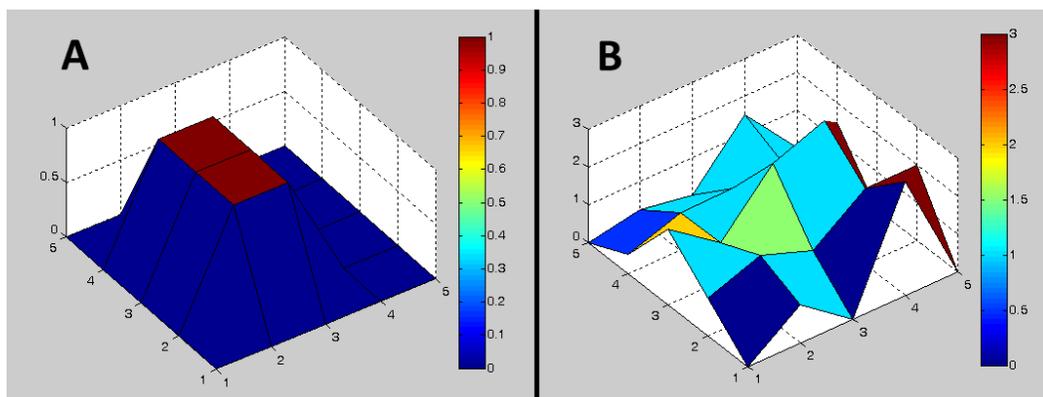

**Figure 4: A distinction between smooth (A) and rugged (B) fitness landscapes. Notice that ruggedness can be defined as a greater amount of fitness variation across a broader range of potential environments.**

*2.2.1 Three generalized properties of adaptability.* There are three general properties of augmented systems that relate to its adaptability. The first is the generalized degree of evolvability. This can be defined as the capacity to meet the challenges provided by the environment by changing some attribute of the physiology. This is also a definition of a variable called $E_d$. The second of these is the degree of robustness, which can be defined as the capacity to meet the challenges provided by the environment within the current boundaries of the physiological system. This is also a definition of a variable called $R_d$. The third of these is the degree of brittleness, which translates into no capacity to meet the challenges of the environment. This is also a definition of a variable called $B_d$. As with fitness, brittleness is also highly contextual: being "brittle" in one setting does not translate into universal brittleness.

*2.2.2 Phase portrait description of states.* Each of these variables can be mapped to a set of systemic states that can provide a quick assessment of the system when selection is compared against fitness[6]. Figure 5 shows this phase portrait representation.

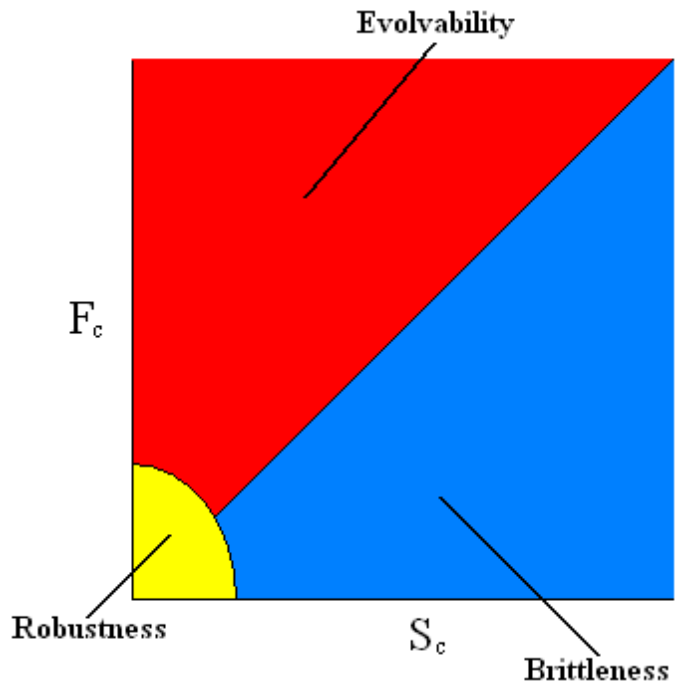

**Figure 5: Phase portrait representation of different states in relation to fitness and selection variables. While the regions representing evolvability (red) and brittleness (blue) are generally stable across contexts, the region representing robustness (yellow) can expand radially away from 0, 0 in particular contexts.**

*2.2.3 Adaptability expressed as comparative functions.* Fitness and selection can also be represented as two continuous functions (see Figure 7). These continuous functions can be analyzed using sets of differential equations that can be compared to one another. In the first example, learning can be defined as environmental sampling. During environmental sampling, the physiological system undergoes a gradient descent on the fitness landscape. Under these conditions, *dS / dt* is larger than the value for *dF / dt*, where the value of the selection function exceeds that of the fitness function. In the second example, memory consolidation occurs subsequent to environmental sampling, and this results in the physiological undergoing a gradient ascent on the fitness landscape. Under these conditions, *dS / dt < dF/ dt*, where the value of the fitness function exceeds that of the selection function.

---

[6] **Notes:** In the case of robustness: $S_c / F_c \sim 0$ (see Figure 5). In the case of evolvability: $F_c / S_c > 0$ (see Figure 5). In the case of brittleness: $S_c / F_c > 0$ (see Figure 5).

*2.2.4 Equilibrium point between selection and adaptation.* Between these two examples, there exists an equilibrium point defined as *dS / dt = dF/ dt*, where the selection and fitness functions are equal. This equilibrium point is a general indicator of robustness in the relationship between the equilibrium point and time. When *t* is much greater than 0, the system is less robust, while *t* values greater than or equal to 0 predict a more robust system. This result indicates that when system challenged with environmental selection, physiological processes and their indicators can respond quickly. In these cases, the minima on the fitness landscape are relatively shallow. This also suggests that the physiological system needs to perform less sampling during times of environmental challenge (See Figure 7).

## 3.0 Learning and Memory Example

Let us suppose an adaptive behavior and physiological response state gauge that can provide an assessment of fitness using a single parameter. This parameter has a discrete output that can range from 0 to 9, and constitutes a toy problem (see Figure 6 and Table 1). This problem is the basis for a four-step model of augmented learning, which is described in Section 3.1.

## 3.1 Four-step adaptability-based model of augmented learning

In the first step, performance on an activity such as driving in daylight, dry conditions and moderate traffic is achieved at a certain level. During this same step, the physiological state gauge reading is "6". In the second step, environmental conditions are degraded such as driving in cloudy conditions in the mountains. The physiological state gauge reading is "3". During this same step, performance is likewise degraded, and the physiological state gauge reading reduced by "3". In the third step, learning has occurred and performance increases on both activities. During this same step, the physiological state gauge reading increases by "5" to a final reading of "8" for both activities (averaged together, small standard deviation). Overall, there are specific interactions between environmental selection presented by activity and fitness exhibited in response to environmental challenge (see Figure 6 and Table 1).

## 3.2 What happens computationally and physiologically?

There are several potential mechanisms behind the adaptability-based model. One possibility is "no free lunch" learning (see Stone, 2007). The sampling function characterized by new environmental information and expected degree of selection function represented by previously consolidated memory and the ability of physiological processes to produce an adaptive behavioral response either align isomorphically[8] or do not. The degree of alignment (e.g. isomorphy between individual fitness and environmental challenges) determines ease of learning and the degree of adaptability achieved.

---

[8] isomorphy can be defined as an equivalent mapping between two discrete mathematical sets, or from one context to another.

The second possibility is the existence of an adaptive ratchet mechanism (for description of mechanism in physical systems, see Freund and Poschel, 2000): a perturbation of the system in the form of environmental selection pushes capacity of system towards otherwise hidden states. This induction of states not normally subject to expression nor selection induces an adaptive response.

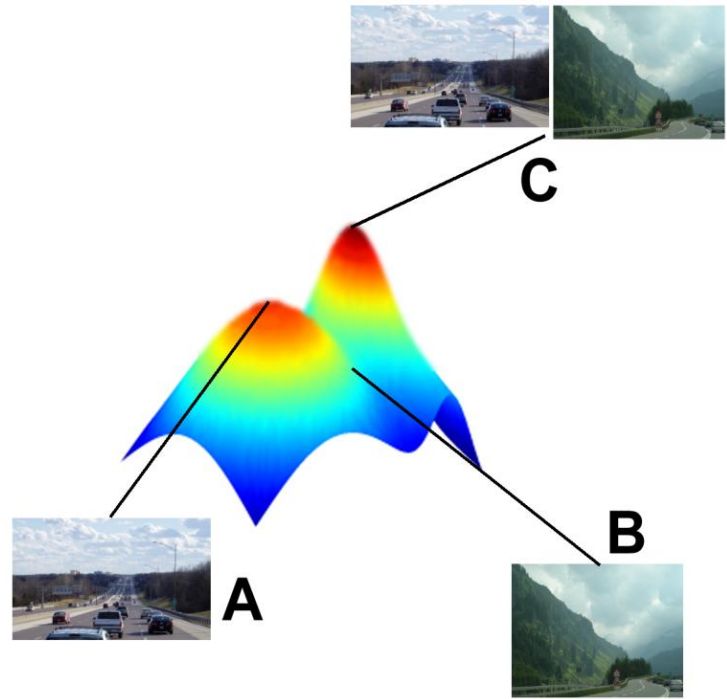

**Figure 6. Graphical representation of learning and memory example. A: gauge reading of "6", B: gauge reading of "3", C: gauge reading of "8".**

**Table 1. Quick description of adaptability-based model as laid out in Figure 6**

| **Let's look at what happens on the fitness landscape:** |
|---|
| * organism encounters a challenge, must adapt (learning and memory). |
| * to adapt, organism must sample new environment. |
| * sampling drives organism into a fitness landscape local minimum. |
| * consolidating old learning with new allows organism to make a gradient ascent. |

The third possibility is that mitigation can be achieved by working with adaptive mechanisms instead of parallel to them. In this case, optimal behavior can emerge from a wide range of activities and environmental conditions. In particular, this mitigation strategy is designed for a generalized stress response that is manifest from social behavior to mechanisms at the genomic scale (Sapolsky, 2004; Gasch *et.al*, 2008). At the molecular and cellular level, the generalized response to exercise and inflammation work in a similar way. In this regard, learning and memory as a process analogous to those that combat effects of inflammation or aging. Connections have been found between such stimulation and cell survival or enhanced function, and may allow us to better understand how molecular and physiological regulatory processes contribute to a complex behavioral phenotype.

### 3.3 The four-step adaptability model as it relates to physiological regulation

Physiological regulation drives this system at multiple scales. One of these is learning and memory at the molecular scale. Learning and memory, or rather it cellular-level correlate long-term sensitization, is controlled by a number of adaptive processes at different scales. When a memory is consolidated, long-term sensitization is established between many synaptic connections throughout brain regions such as the hippocampus (Anderson et.al, 2007) for declarative memory and the insular/amygdalar pathway for fear conditioning (Bermudez-Rattoni, 2004). This occurs via the MAPK-ERK signal transduction pathway, which is regulated by a balance of biomolecules called phosphotases and kinases. Equilibrium between these two biomolecular sources play a role in controlling the biological activity of proteins that modulate gene expression (Sweatt, 2003), which in turn determines the state and extent of memory consolidation. Endocrine fluctuations over time can also affect the forementioned processes, which produce the optimal behavioral output in question. The regulation of all of these physiological processes may also vary across individuals, different physiological processes, and environmental contexts.

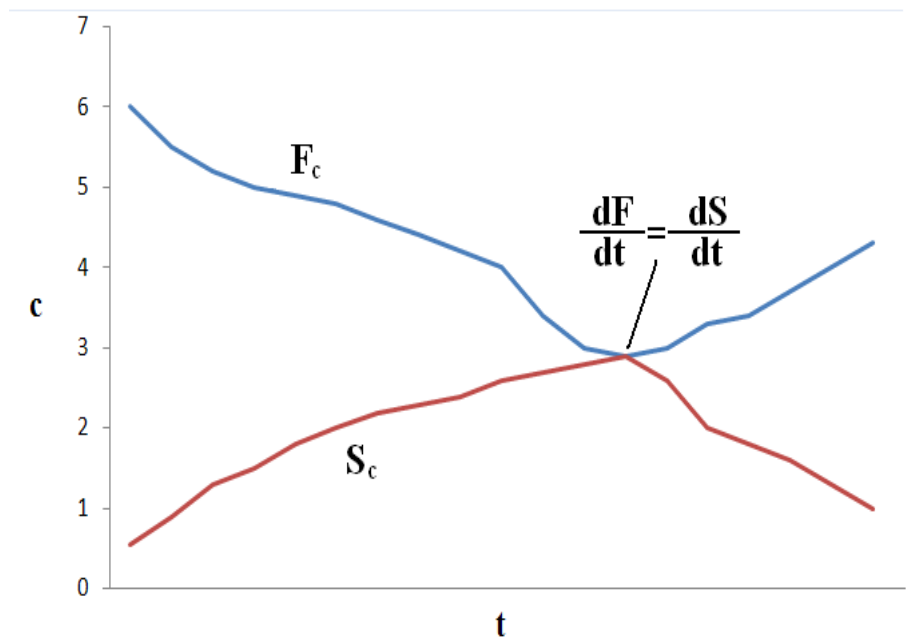

**Figure 7. Graphical representation of selection and fitness dynamics during environmental sampling and memory consolidation. The equilibrium point is shown to be where $F_c$ and $S_c$ intersect.**

## 3.4 Examples from performance and behavior

One operational example of how an individual descends and then ascends the surface of a fitness landscape during performance can be found in interference between different types of memory consolidation in response to a complex stimulus. Some forms of memory consolidation such as fear conditioning are quick (on the order of a single presentation). Other forms of memory, such as declarative memory, involve different physiological subsystems at the scale of neural circuitry which require on the order of tens of repeated presentations. In many contexts, two or more subsystems such as these must work together during the learning of a complex stimulus in a manner similar to what was outlined for the four-step model. In the case of fear conditioning and declarative memory, a fear memory acquired in one time step can suppress the acquisition of a declarative memory. However, the ability to store and recall both fear and declarative memories within the same context may improve the overall fitness of an individual in the long-term.

*3.4.1 Hill-climbing vs. the power law of practice.* When an individual or population descends or climbs the surface of a fitness landscape, it is often referred to as hill-climbing. When applied to learning-related physiological adaptation, hill-climbing must be reconciled with the power law of practice (Newell, 1990). At first glance, these two models have contradictory elements. However, they can be reconciled by considering systems in equilibrium versus systems under dynamical selection. The basic idea of the power law of practice is that decreases in a parameter such as reaction time are steeper initially, and converge upon an optimal value once learning has ended and memory has consolidated. However, it is important to note that power-law behavior may only occur in systems that are analyzed in equilibrium, with a stimulus that is held constant. By contrast, the performance landscape model allows for a continually changing stimulus.

## 4.0 How this is a better mitigation strategy

There are several reasons why this approach might be an improvement on currently utilized models. A complex adaptive system may be able to tailor individual mitigation strategies, as every individual will respond differently but respond within the parameters of this model. Secondly, this model is not overly simplistic. An adage of the complexity theory community is the phrase "keep it complex, stupid!" Indeed, sometime Occam's razor is an incomplete assumption (Sober, 1981).

A more operational reason is that such a model allows the control system to sample a greater range of a person's "learning curve" as defined by physiological adaptation. This is important, because there are multiple forms of learning and memory. As a theoretical construct, this model covers the processes behind many of these in terms of basic functionality. It may be able to deal with mitigation in both a contextually-specific and general manner.

Future directions involve developing combinatoric algorithms that capture the adaptability of specific individuals for specific activities to trigger mitigation. To move in this direction, consider that this model allows for neutral[9] scenarios to be optimal, which means that many possible pathways can be taken to achieve optimal behavior.

---

[9] Neutral evolution can be defined as adaptation along a number of pathways with an equivocal fitness. Neutral evolution is combinatoric in sense that $n$-choose $m$ scenarios are required to get there.